\documentstyle[prl,aps,epsf,psfig]{revtex}
\begin{document}
\twocolumn[\hsize\textwidth\columnwidth\hsize
           \csname @twocolumnfalse\endcsname
\title{Hall voltage sign reversal  in type II
superconductors}
% ========== authors ==========d
\author{Jan Kol\'a\v{c}ek, Petr Va\v{s}ek}
\address{Institute of Physics, ASCR, Cukrovarnick\'a 10,16253 Prague 6,
Czech Republic}
\maketitle
\begin{abstract}
% ========= abstract =============
The Hall voltage sign reversal is consistently explained by the model
in which vortices  with the superconducting and normal state charge
carriers are regarded as three  subsystems mutually connected
by interactions. The equations of motion for these three subsystems are
solved simultaneously and a new formula for the  Hall resistivity
in the flux flow regime is obtained.
It is shown that  it is possible to explain qualitatively
experimental data by this model.
\end{abstract}
% ------------ PACS numbers --------
\vspace{5mm}
%Corresponding author : Jan Kol\'a\v{c}ek, e-mail : kolacek@fzu.cz,
%      fax : 4202 3123184
    \vskip2pc]

\section{Introduction}

The dynamics of vortices shows a rich behaviour that is still not completely
understood. There is a general belief that moving vortices
significantly contribute to the Hall voltage sign reversal
in the mixed state of superconductors which is
one of the most intriguing and controversial transport phenomena. One of the
earliest attempts to explain this Hall anomaly was made using
Bardeen and Stephen
\cite{65Bardeen}  or Nozieres and  Vinnen \cite{66Nozieres} models. However
these theories are unable to provide a satisfactory explanation
of experimental data.

Starting from the fact that vortices moving with velocity ${\bf v}_L$ create
electric field
$ {\bf E}_v = - {\bf v}_L \times {\bf B}_v $
(${\bf B}_v $ is  magnetic field carried by vortex)
Hagen at all \cite{90Hagen} argued that vortices
moving antiparallel to the transport current must be responsible for the Hall
voltage sign reversal. Several theories were suggested  to explain this
behavior. They took into account
particle hole asymmetry \cite{71Fukuyama,97Nishio},
the presence of antivortices \cite{92Jensen,94Viret} or
vacancies in a pinned vortex lattice \cite{97Ao}.
Wang et al. \cite{WDT} developed a theory explaining
observed results by pinning and thermal
fluctuation effects. Recently Zhu et al. \cite{ZXWZZ} extended Wang's theory
taking into account also vortex-vortex interaction.
Effect of pinning on the vortex Hall resistivity was also dealt in \cite{KV}.
However, measurements made on samples with different correlated disorder
\cite{98Beam} confirmed that sign reversal is an intrinsic property of type II
superconductors and also other experiments
did not confirm the role of pinning in Hall
voltage change of sign \cite{BLC,LGG}.
Also measurements at high current densities \cite{G99} suppressing the
effective pinning show only little effect.
Attention was devoted also to the attempt to explain the Hall anomaly
by taking into account the charge of  the vortex core \cite{95Feigelman,95Khomski}.
As the chemical potential differs in the normal and the superconducting
phases,
the vortex core may become charged. Simple physical considerations
show that the charge of the vortex core is opposite to the sign of the
dominant charge carriers \cite{95Khomski} and it was recently experimentally
verified experimentally by Matsuda and Kumagai \cite{99Matsuda}.
Feigelman's \cite{95Feigelman} formula can be used to describe
the experimental data quite well (see e.g. \cite{97Martin,98Nakao}),
but to achieve it the carrier density in the core must be supposed
to be larger than outside, which seems to be incorrect. On the other hand
Khomskii \cite{95Khomski} supposes the correct sign of the vortex
core charge, but Feigelman \cite{95Feigelman} argues that the
additional transverse force used by Khomskii has the opposite sign,
which contradicts the result for the Magnus force in the
Galilean invariant case. Moreover Wang et.al. argue that in the low
magnetic field limit both these models lead to H-dependence of Hall angle,
which violates their experimental finding obtained
in ultra low magnetic field \cite{98Wang}.
Currently Hall anomaly is still considered to be an unsolved
problem and new models are proposed (see e.g. the mixed charge model
proposed by Ji and Wang \cite{99Ji}).

 For correct understanding of the vortex dynamics it is necessary to know
the forces acting on the vortices. Using a very general Berry phase arguments
 Ao and Thoules \cite{93Ao} concluded that Magnus force is the only transverse
 force on the vortex. If this is true, no transverse force from quasiparticles
 and impurities would act on the  vortex. Recently Sonin
\cite{96Sonin,97Sonin} doubted
this result arguing that the effective Magnus force, where Iordanskii force
from quasiparticles \cite{66Iordanskii} and Kopnin-Kravtsov force from
impurities \cite{76Kopnin} are included
should be taken into account. Krasnov and Logvenov
\cite{97Krasnov} have used
effective Magnus force proposed by Sonin for calculation of
transport properties in the presence of transport current and
temperature gradient. In the discussion of their results they
omitted the Iordanskii force and moreover they did not take into
account reaction force of vortices on superconducting and normal
particles.

In our model we suppose that the Magnus force is responsible for the
 interaction between vortices and superconducting fluid, while Lorentz force
(which is equivalent to the Iordanskii force) is responsible for the
interaction between vortices and the normal state fluid. In our approach we
treat the vortices, superconducting and normal state fluid as three mutually
interacting subsystems and simultaneously  solve their equations of motion.
The solution quite naturally explains the Hall voltage sign reversal and
simple numerical calculations showed that the model is able to explain at
least qualitatively the experimental data. Moreover it is shown
that Hall anomaly exists in the flux flow regime, i.e. it is not necessary
to take into account the pinning.

\section{The  model}
%====================================
Vortices, superconducting and normal state fluids form three mutually
interacting subsystems, consequently their equations of motion must be solved
simultaneously. At the beginning let us briefly summarize the forces felt by the
 vortices
(v), superconducting (s) and normal state (n) charge carriers.
Vortices interact with superconducting charge carriers by the Magnus force. If
a vortex oriented along the z-axis moves with velocity ${\bf v}_L$
then it feels the force
%---------Magnus vortex----------
\begin{eqnarray} \label{eqmagnusv}
{\bf F}_{M} (v)
    &=& {e \over |e|} {{n_s h} \over 2} ({\bf v}_s - {\bf v}_L ) 
         \times {\bf z} \nonumber \\ 
    &=& {e \over |e|}  m _{v} f_s \Omega ({\bf v}_s - {\bf v}_L ) 
         \times {\bf z},
\end{eqnarray}
where $n_s = f_s n$ is the density of superconducting charge carriers having
 electric
charge $e$ and velocity ${\bf v}_s$ . Vortices are supposed to behave like
quasiparticles;
$m_v$ denotes vortex mass per unit length. If the superconducting fluid does
 not move the
Magnus force is perpendicular to the vortex velocity so that in absence of
 damping the
free vortex makes circular motion with angular frequency
 $ \Omega = {{n h} / {2 m_v}} $.
The superconducting charge carriers must feel the reaction force
% -------Magnus supra--------
\begin{eqnarray} \label{eqmagnuss}
{\bf F}_{M} (s)
   = - {{n_v} \over {n_s}} {\bf F}_{M} (v)
   = -{e \over |e|}  m \omega _c ({\bf v}_s - {\bf v}_L ) \times {\bf z},
\end{eqnarray}
where $n_v$ is the vortex density and $ \omega_c=eB/m$
is the cyclotron frequency of
superconducting charge carriers having the effective mass $m$ in the magnetic
field
${\bf B} = n_v \Phi_0 {\bf z}$, which is the averaged field inside the
 superconductor.

Let us consider the vortex system, in which the mean distance between
vortices
is small in comparison with the penetration depth. In this case the magnetic
 field in the
superconductor is almost homogeneous and the normal state charge carriers
 moving
with the velocity ${\bf v}_n$  must feel the Lorentz force. Using Aharonov-Casher
lagrangian \cite{Aha} it can be shown that also in this case
relative velocity ${\bf v}_n - {\bf v}_L $ is decisive, so that the force
felt by the superconducting charge carriers can be
written in  following form

%-------Lorentz normal-----------
\begin{eqnarray} \label{eqLorentzn}
{\bf F}_{L} (n)
   = e ({\bf v}_n - {\bf v}_L ) \times {\bf B}
   = {e \over |e|} m \omega _c ({\bf v}_n - {\bf v}_L ) \times {\bf z}.
\end{eqnarray}
Vortices carrying the magnetic field  act on the normal state fluid,
and consequently they must feel the reaction force
%-------Lorentz vortex------------
\begin{eqnarray} \label{eqLorentzv}
{\bf F}_{L} (v)
     = - {{n_n} \over {n_v}} {\bf F}_{L} (n)
     &=& - {e \over |e|} f_n { {nh} \over 2} ({\bf v}_n - {\bf v}_L ) \times
 {\bf z} \nonumber \\
     &=& - {e \over |e|} m_v f_n \Omega ({\bf v}_n - {\bf v}_L ) \times {\bf z}.
\end{eqnarray}
To simplify the formulae in the following text the electric charge $e$ is
 supposed to be
positive. Vortex pinning and damping will be characterized by the vortex
pinning
frequency $\alpha = \sqrt{\kappa /m_v}$
($\kappa$  is the commonly used pinning constant) and vortex relaxation
time $\tau _v$, respectively. The normal state fluid damping will be characterized by the
normal state fluid
relaxation time $\tau _n$.
In the Hall effect experiments the electric current ${\bf j}=j_x {\bf x}$ is
controlled
experimentally and the velocities ${\bf v}_L$, ${\bf v}_s$, ${\bf v}_n$
together with the electric field $\bf E$  must be
determined from the  expression for current
%---------- current ---------
\begin{eqnarray} \label{current}
{\bf j} &=& n e (f_s {\bf v}_s +f_n {\bf v}_n )
\end{eqnarray}
and from the following three equations of motion
\begin{eqnarray} \label{eqsmotion}
{\dot {\bf v}}_s &=& {e \over m} {\bf E} -
       \omega _c ({\bf v}_s - {\bf v}_L) \times {\bf z}
\end{eqnarray}
\begin{eqnarray}
{\dot {\bf v}}_n &=& {e \over m} {\bf E} +
       \omega _c ({\bf v}_s - {\bf v}_L) \times {\bf z} -
         {1 \over \tau _n} {\bf v}_n
\end{eqnarray}
\begin{eqnarray}
{\dot {\bf v}}_L = - \alpha^2 {\bf r}_L - {1 \over \tau _v} {\bf v}_L
      &+& f_s \Omega ({\bf v}_s - {\bf v}_L) \times {\bf z} \nonumber \\
      &-& f_n \Omega ({\bf v}_n - {\bf v}_L) \times {\bf z}
\end{eqnarray}

The same set of equations has been recently used successfully for
the study of high frequency vortex dynamics and for
interpretation of the far infrared magnetoconductivity \cite{kol}.

In the steady state the accelerations on the left hand sides of the
equations of
motion are equal to zero. The solution depends on the magnitude of the current
density used
for the measurement.
For high currents Magnus force exceeds the vortex pinning, vortices
move and in the first approximation the effect of vortex pinning can be
neglected.
After substituting $\alpha = 0$ to equation of motion 
(\ref{eqsmotion}) the vortex velocity is
%---------- velocity_zp
\begin{eqnarray} \label{velocity_zp}
v_{Lx} &=& [(\tau_n^2 \omega_c^2 + f_s^2(f_s - f_n))\tau_v \Omega
    + 2 \tau_n \omega_c f_n f_s]
     {\beta j_x } \\
v_{Ly} &=& -[2\tau_n \omega_c \tau_v \Omega f_n f_s
     + \tau_n^2 \omega_c^2 (f_s - f_n) - f_s^2]
     {\beta j_x }
\end{eqnarray}
where $\beta = {\tau_v \Omega}/{D n e}$ and
%------------ denominator
\begin{eqnarray} \label{denominator}
D &=& [\tau_v^2 \Omega^2 +(f_s - f_n)^2]\tau_n^2 \omega_c^2
   + 8 f_s^2 f_n \tau_n \omega_c \tau_v \Omega  \nonumber \\
   &+& f_s^2 (f_s - f_n)^2 \tau_v^2 \Omega^2 + f_s^2 .
\end{eqnarray}
The total current in the x-axis direction $j_x$ is the sum of the
superconducting current
%------------- jsx_zp
\begin{eqnarray} \label{jsx_zp}
j_{sx} &=& [(f_s (f_s - f_n)^2 + \tau_n^2 \omega_c^2) \tau_v^2 \Omega^2
    + 6 f_n f_s \tau_n \omega_c \tau_v \Omega \nonumber \\
    &+& \tau_n^2 \omega_c^2 (f_s - f_n) + f_s] {f_s \over D} j_x
\end{eqnarray}
and the normal state current
%------------- jnx_zp
\begin{eqnarray} \label{jnx_zp}
j_{nx} = [\tau_n^2 \omega_c^2 \tau_v^2 \Omega^2
    + 2 f_s^2 \tau_n \omega_c \tau_v \Omega \nonumber \\
    - \tau_n^2 \omega_c^2 (f_s - f_n)] {f_n \over D} j_x ,
\end{eqnarray}
while the superconducting current in the y-axis direction is cancelled by
 the normal
fluid backflow current
%------------- jsy_zp
\begin{eqnarray} \label{jsy_zp}
j_{sy} = - j_{ny} 
     =  [-\tau_n \omega_c \tau_v^2 \Omega^2 (f_s - f_n) \nonumber \\
    + 2 \tau_n^2 \omega_c^2 \tau_v \Omega + \tau_n \omega_c]
    {f_n f_s\over D} j_x ,
\end{eqnarray}
The longitudinal and Hall resistivities may be  expressed as
%-------- rho_zp
\begin{eqnarray} \label{rho_zp}
\rho_{xx} &=& {E_x \over j_x}
     = [\tau_n \omega_c f_n (1 + \tau_v^2 \Omega^2) \nonumber \\
     &+& \tau_v \Omega (f_s^2 + \tau_n^2 \omega_c^2 )]
       {\omega_c \over \epsilon_0 \omega_p^2 D},\\
\rho_{xy} &=& {E_y \over j_x}
     = [(f_s - f_n) ( f_s f_n \tau_v^2 \Omega^2 - \tau_n^2
\omega_c^2)   \nonumber \\
       &-& f_s(1+4f_n \tau_n \omega_c \tau_v \Omega)]
       {\omega_c \over \epsilon_0 \omega_p^2 D}.
\end{eqnarray}
The last formula quite naturally explains the change of sign in the Hall
effect
measurements. At zero temperature where $f_n = 0$, the Hall
resistivity
$\rho_{xy}
  = - \omega_c / \epsilon_0 \omega_p^2 (1 + \tau_v^2\Omega^2)$
is negative, while for temperatures above $T_c$ it has the
positive value $\rho_{xx} = \omega_c / \epsilon_0 \omega_p^2 $.
The conductivity may be expressed as
%------------ sig_zp
\begin{eqnarray} \label{sig_zp}
\sigma_{xx} &=& {\epsilon_0 \omega_p^2 \over \omega_c \gamma}
       [f_n \tau_n \omega_c (1 + \tau_v^2 \Omega^2)
       + \tau_v \Omega (f_s^2 + \tau_n^2 \omega_c^2)] ,\\
\sigma_{xy} &=& {\epsilon_0 \omega_p^2 \over \omega_c \gamma}
       [(f_s - f_n) (\tau_n^2 \omega_c^2
        - f_s f_n \tau_v^2 \Omega^2)  \nonumber \\
        &+& 4 f_s f_n \tau_n \omega_c \tau_v \Omega + f_s]
\end{eqnarray}
where $\gamma= 1 + (\tau_n \omega_c + f_n \tau_v \Omega)^2 $.
It is interesting to note, that using $\omega_c = e B/m$ for low magnetic
field it is possible to approximate the Hall
conductivity  as
%------------ aproxsig
\begin{eqnarray} \label{aproxsig}
\sigma_{xy} = A_{-1}/B + A_0 + A_1 B .
\end{eqnarray}
In the low temperature region, where the dependence of normal state fraction
$f_n$  on the
magnetic field is negligible the coefficients can be written as
%------------ AAA
\begin{eqnarray} \label{AAA}
&A&_{-1}= \epsilon_0 \omega_p^2 {m \over e}
      {f_s^2 \tau_v \Omega)
      \over 1 + f_n^2 \tau_v^2 \Omega^2}  \\
%\end{eqnarray}
%\begin{eqnarray}
&A&_{0}= \epsilon_0 \omega_p^2 f_n \tau_n
      {(1 + f_n^2 \tau_v^2 \Omega^2)^2 - f_s^2 \tau_v^2 \Omega^2
      \over (1 + f_n^2 \tau_v^2 \Omega^2)^2 } \\
%\end{eqnarray}
%\begin{eqnarray}
&A&_{1}= {\epsilon_0 \omega_p^2 {e \over m} \tau_v \Omega)\tau_v \Omega
   \tau_n^2}  \nonumber \\
 &\times&{[(1+f_n^2 \tau_v^2 \Omega^2 )^2
      - f_n^2 f_s (f_s +4)\tau_v^2 \Omega^2
      - f_s (1+3 f_n)]
      \over (1 + f_n^2 \tau_v^2 \Omega^2)^3}.
\end{eqnarray}

\section{Results}
The above presented formulae should be applicable mainly to high temperature
superconductors, because in these materials the London penetration length is
rather large, 
\begin{figure}[h]  % F1
\centerline{\parbox[t]{9cm}{
\psfig{figure=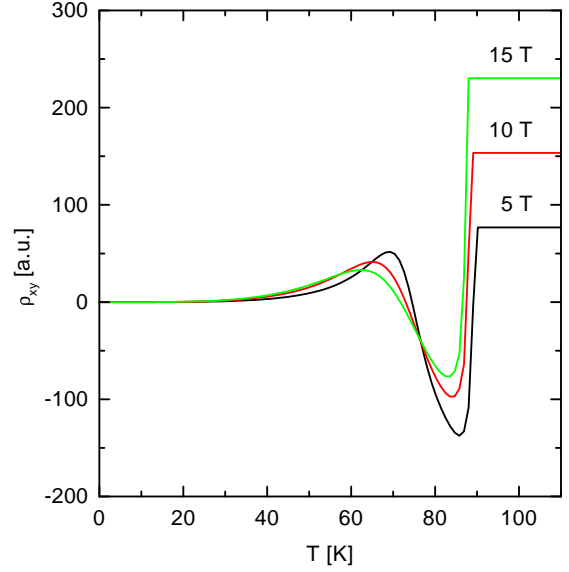,width=9cm,height=9cm}}}
\caption{ Temperature dependence of the Hall resistivity for  
different magnetic fields. Parameters of the model are 
  $\Omega_0 \tau_v = 50, \alpha=0, 1/\tau_n=35 \text{cm}^{-1}$.}
\label{Fig.1}
\end{figure}
so that even in moderate fields the magnetic field 
in the superconductors is almost homogeneous. The coherence length 
is small, so that the redistribution of charge and current density 
in the normal state core may be neglected. To determine the
theoretical dependence of the  Hall resistance on the temperature
and magnetic field, it is necessary to estimate the dependence 
of the parameters involved. 
We suppose that $f_n = (T/T_c)^4$ where the critical temperature $T_c$ 
depends on the magnetic field as $T_c = T_{c0} \sqrt {1-B/B_{c2}} $.
Here $T_{c0}$ and $B_{c2}$ are the critical temperature at zero magnetic field
and critical magnetic field at zero temperature, respectively.  Using the
Hsu's expression for the
vortex mass $m_v = (\pi^2/4) m k_F^2 \xi^2$  \cite{93Hsu},
the  expression for the Fermi wave vector $k_F^2=2 \pi n$  and
$ \xi = {{\hbar v_F} / {\pi \Delta}}$   for the coherence length,
it is possible to show that
$\Omega = \Delta^2/E_F$. Supposing that the Fermi energy $E_F$ is a constant,
while the gap depends on the temperature and magnetic field approximately as
$\Delta = \Delta_0 \sqrt{\text{cos}((\pi T^2/2T_c^2)}$
\cite{89Popel},  we will take
$\Omega = \Omega_0 \text{cos}(\pi T^2 /2T_c^2)$.
The cyclotron frequency is proportional to the magnetic field
$\omega_c = e B/m$, and the
relaxation times $\tau_n$, $\tau_v$ are supposed to be constant
 in the first approximation. 
\begin{figure}[h]  % F2
\centerline{\parbox[t]{8cm}{
\psfig{figure=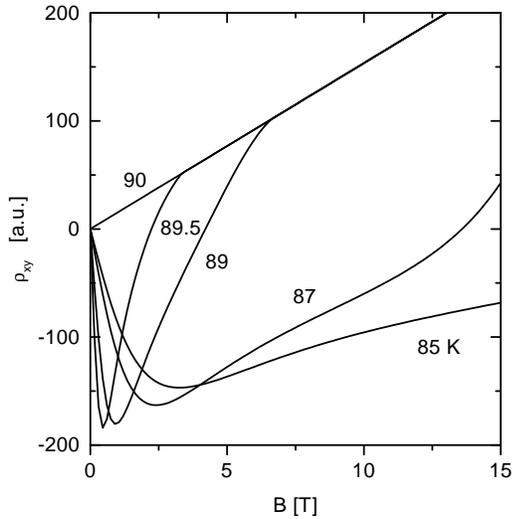,width=8cm,height=8cm}}}
\caption{Magnetic field dependence of the Hall resistivity for
different temperatures. The parameters are same as in the Fig.1}
\label{Fig. 2}
\end{figure}
Using above mentioned expressions
the theoretical temperature and magnetic field dependencies of the Hall
resistivity in the zero pinning regime (eq. 16) plotted in the Fig. 1, 2
 reproduce qualitatively
the data observed experimentally by many authors.  The second sign
 change from negative to
positive value  is clearly
seen  from the Fig.1 . Such behavior is observed experimentally in Bi ,
Tl and Hg based superconductors. Moreover detailed analysis of
formula (16) reveals  a third sign change at even lower
temperatures. Such change of sign was recently observed by Kang
et al.
\cite{Kang} on HgBaCaCuO. However we should note that it is
questionable if such
experimentally observed third sign change is not a consequence  of
the presence of
another superconducting phase as it was suggested in \cite{vasek}.

\section{Conclusion}

We have explained the Hall voltage sign reversal in the
superconductors below the critical temperature
considering vortices, superconducting and normal
state fluids as mutually interacting subsystems where also
reaction forces are taken into account. Results of calculations
reveal up to three changes of sign of Hall resistivity
with decreasing temperature.
 The experimentally observed
dependence of the Hall conductivity  on magnetic field
\cite{95Ginsberg,Kim} also confirms our results (eq. 19),where a
 term independent on magnetic field is added in contrast to the
previous theories \cite{Dorsey,Kopnin}.
Without any
special assumption about pinning strength or fluctuations as was made by
Wang et al \cite{94Wang} the theoretical curves qualitatively explain
 experimental  results of the Hall effect of  superconductors in
the mixed state. 
It is worth noting that in the above presented theory 
the boundary conditions on the sample surface are not considered. 
It seems clear that for quantitative description of 
the experimental data finite dimensions of the measured sample 
should be taken into account.
\acknowledgments

%=========================
This work was supported by M\v{S}MT project Kontakt ME 160 
and GACR $\sharp106/99/1441$,  
GACR $\sharp104/99/1440$, GAAV A$1010919$ grants .

%=========================


\begin{thebibliography}{99}
\bibitem{65Bardeen} J. Bardeen and M. J. Stephen, Phys. Rev. B 49,
1197 (1965)\
\bibitem{66Nozieres} P. Nozieres and W. F. Vinnen, Phil. Mag. 14,
 667 (1966)\
\bibitem{90Hagen} S. J. Hagen, C. J. Lobb, R. L. Green, M. G.Forrester
 and J. Talvacchio, Phys. Rev. B 42, 6777 (1990)\
\bibitem{71Fukuyama} H.Fukuyama, H.Ebisawa, T.Tsuzuki Prog.Theor.Phys. 46, 1028  (1971)\
\bibitem{97Nishio} T.Nishio, H.Ebisawa, Physica C 290, 43 (1997)\
\bibitem{92Jensen} H.J. Jensen, P. Minnhagen, E. Sonin and H. Weber,
Europhysics Lett. 20, 463 (1992)\
\bibitem{94Viret} M. Viret and J. M. D. Coey, Phys. Rev. B 49,
 3457 (1994)\
\bibitem{97Ao} P. Ao, J.Phys.Condens.Matter 10, L667 (1998)\
\bibitem{WDT} Z.D.Wang, J. Dong and C.S. Ting, Phys. Rev. Lett.
72, 3875 (1994)\
\bibitem{ZXWZZ} B. Y. Zhu, D. Y. Xing, Z. D. Wang, B. R. Zhao and
Z. X. Zhao, Phys. Rev. B 60, 3080 (1999)\
\bibitem{KV} N. B. Kopnin and V. M. Vinokur, Phys.Rev.Lett. 83, 4864,
(1999)\
\bibitem{98Beam} D.A.Beam, N.-C.Yeh, F.Holtzberg,
       J.Phys.: Condens. Matter 10, 5955 (1998)\
\bibitem{BLC} R. C. Budhani, S. H. Liou and Z.X. Cai, Phys. Rev. Lett.
71, 621 (1993)\
\bibitem{LGG} W. Leibich, P. Gehringer, W. G\"{o}b, W. Lang, K. Bierleutgeb
and D. B\"{a}uerle, Physica C 282-287, 2321 (1997)\
\bibitem{G99} W. G\"{o}b, W. Lang, J. D. Pedarning, R. R\"{o}ssler and D.
B\"{a}uerle, Physica C317-318, 627 (1999)\
\bibitem{95Feigelman} M. V. Feigelman , V.B. Geshkeibein, A. V. Larkin
and V. M. Vinokur, JETP Lett 62, 834 (1995)\
\bibitem{95Khomski} D. I. Khomski and A. Freimuth, Phys. Rev. Lett.
75,  1384 (1995)\
\bibitem{99Matsuda} Y. Matsuda, K.Kumagai , Proc. of LT22,
Finland, 398 (1999)\
\bibitem{97Martin} J.I.Martin, M.Velez, F.Guinea, J.L.Vicent,
    Phys.Rev. B 55, 5659 (1997)\
\bibitem{98Nakao} K.Nakao, K.Hayash, T.Utagawa, Y.Enomoto, N.Koshizuka,
      Phys.Rev.B 57, 8662 (1998)\
\bibitem{98Wang} Wang Yun-Ping, Zhang Dian-Lin, Zhao Bai-Ru,
      Modern Phys.Lett.B 12, 719 (1998)\
\bibitem{99Ji} H.L.Ji, K.W.Wong, Phys.Lett.A 256, 66 (1999)\
\bibitem{93Ao}  P. Ao and D. J. Thouless, Phys. Rev. Lett.
70, 2158 (1993)\
\bibitem{96Sonin} E. B. Sonin, Czech. J. Phys. 46, 911 (1996)\
\bibitem{97Sonin} E. B. Sonin, Phys. Rev. B 55, 485 (1997)\
\bibitem{66Iordanskii} S. V. Iordanskii, Sov. Phys. JETP 22,
160 (1966)\
\bibitem{76Kopnin} N. B. Kopnin and V. E. Kravtsov, Sov. Phys. JETP 44,
 861 (1976)\
\bibitem{97Krasnov} V. M. Krasnov and G. Yu. Logvenov, Physica C
 274, 268 (1997)\
\bibitem{Aha} Y. Aharonov and A. Casher, Phys. Rev. Lett.
53, 319 (1984)\
\bibitem{kol} J. Kol\'{a}\v{c}ek and E. Kawate, Phys. Lett.  A 260, 300
(1999)\
\bibitem{93Hsu} T. C. Hsu, Physica C 213, 305 (1993)\
\bibitem{89Popel} R. P\"{o}pel, J. Appl. Phys. 66, 5950
(1989)\
\bibitem{Kang} W. N. Kang, B.W. Kang, Q.Y. Chen,J.Z. Wu, Y. Bai,
W. K. Chu, D. K. Christen , R. Kerchner and S.-I. Lee, Phys.Rev.B 61,
722 (2000)\
\bibitem{vasek} P. Va\v{s}ek, I. Jane\v{c}ek, Int. J. Mod. Phys.
B 13, 3741 (1999)\
\bibitem{95Ginsberg}D. M. Ginsberg and J. T. Manson, Phys. Rev. B 51, 515
(1995)\
\bibitem{Kim}  W.-S. Kim, W.N. Kang, S. J. Oh, M.-S. Kim, Y. Bai,
S.-I. Lee, CH. H. Choi and H.-C. Ri, Physica C324, 77 (1999)\
\bibitem{Dorsey} A. T. Dorsey, Phys. Rev B46, 8376 (1992)\
\bibitem{Kopnin} N. B. Kopnin, B. I. Ivlev, V. A. Kalatsky, J.
Low. Temp. Phys. 90 , 1 (1993)\
\bibitem{94Wang} Z. D. Wang, J. Dong and C. S. Ting, Phys. Rev.
Lett. 72, 3875 (1994)\
\end{thebibliography}
\end{document}